\documentclass{PoS}

\usepackage{slashed}

\usepackage{amsmath,amsfonts,amssymb}
\usepackage{graphicx}
\usepackage{calligra,fontenc, times, mathptmx}
\usepackage[T1]{fontenc}

\title{Quasi-PDFs and pseudo-PDFs}

\ShortTitle{Quasi-PDFs and pseudo-PDFs}

\author{\speaker{Anatoly Radyushkin}  \\%\thanks{A footnote may follow.}\\
         Physics Department, Old Dominion University, Norfolk,
             Virginia  23529, USA \\ 
Thomas Jefferson National Accelerator Facility,
              Newport News, Virginia  23606, USA \\
        E-mail: \email{radyush@jlab.org}}

\abstract{

We discuss the physical nature of   quasi-PDFs, especially  the  reasons 
for the strong nonperturbative evolution pattern which they reveal in actual lattice gauge calculations.
We argue that quasi-PDFs may be  treated  as hybrids of PDFs and the 
rest-frame 
momentum distributions of partons. 
The latter is also responsible for the transverse momentum dependence of TMDs.
 The  resulting convolution  structure of  quasi-PDFs 
necessitates  using  large probing momenta $p_3 \gtrsim 3$ GeV 
to 
 get reasonably close to the PDF limit.
 To deconvolute the rest-frame distribution effects, 
 we propose   to use  a method based directly on the coordinate representation.
 We treat  matrix elements $M(z_3,p_3)$ as 
  distributions  ${\cal M} (\nu, z_3^2)$  depending on 
  the  Ioffe-time  $\nu = p_3 z_3$ and the distance parameter $z_3^2$. 
  The rest-frame spatial distribution is given by ${\cal M} (0, z_3^2)$.  Using the reduced 
  Ioffe function  
  ${\mathfrak M} (\nu, z_3^2) \equiv  {\cal M} (\nu, z_3^2)/ {\cal M} (0, z_3^2)$ 
  we divide out the rest frame effects,
  including the notorious link renormalization factors. 
 The $\nu$-dependence remains intact and determines the shape of PDFs
   in the small $z_3$ region.  The residual $z_3^2$ dependence of the 
    ${\mathfrak M} (\nu, z_3^2)$ 
    is governed by 
   perturbative evolution.     The Fourier transform of ${\cal  M} (\nu, z_3^2)$ produces 
pseudo-PDFs  ${\cal P}(x, z_3^2)$   that generalize
the light-front PDFs   onto spacelike intervals.  On the basis of these findings we propose 
   a new method for  extraction of PDFs from lattice calculations.

}

\FullConference{QCD Evolution 2017\\
		22-26 May, 2017\\
		Jefferson Lab Newport News, VA - USA\\
		\vspace{10pt}
                $\textnormal{JLAB-THY-17-2601}$}

\begin{document}

\section{Introduction}

The usual parton distribution functions (PDFs) $f(x)$ \cite{Feynman:1969ej}  
measured in deep inelastic scattering and other inclusive processes 
are  defined through  matrix elements
of  certain  bilocal operators on   the light cone $z^2=0$. This fact  prevents 
a direct extraction  of these functions  from Euclidean    lattice 
 gauge  theory  simulations.  
 Still,  recently, X. Ji  \cite{Ji:2013dva} proposed
  to use 
separations $z=(0,0,0,z_3)$ which are purely space-like.
Then one can define parton distributions   in the $k_3=yp_3$ component of the parton momentum.
These   quasi-PDFs  $Q(y,p_3)$ 
approach the light-cone PDFs   $f(y)$ 
     in the limit of large hadron momenta  \mbox{$p_3 \to \infty$}. 
     The qusidistribution  method can be also applied to distribution amplitudes (DAs).  
Lattice calculations of  
quasi-PDFs   were  discussed  
in Refs.  \cite{Lin:2014zya,Chen:2016utp,Alexandrou:2015rja}. The results for 
 the pion quasi-DA   were reported  in Ref. \cite{Zhang:2017bzy}. 
The lattice studies demonstrated a very strong
change of quasi distributions with the probing momentum $p_3$,
which cannot be explained by perturbative evolution. 

In our recent papers \cite{Radyushkin:2016hsy,Radyushkin:2017gjd}, 
we  have demonstrated that  quasi-PDFs can be obtained from 
  the   transverse momentum dependent 
 distributions    (TMDs) ${\cal F} (x, k_\perp^2)$.
 We also showed that the  $k_\perp^2$-dependence of the TMDs
 plays the major role in the 
nonperturbative $p_3$-evolution of
quasi-PDFs and quasi-DAs. In these papers, we have based  our studies on 
 the formalism of  virtuality distribution functions 
 \cite{Radyushkin:2014vla,Radyushkin:2015gpa}.
The TMD/quasi-PDF relation allows to use  simple models for TMDs
for  building 
models for the    nonperturbative evolution of
quasi-PDFs and quasi-DAs.  
The results  obtained in our papers \cite{Radyushkin:2016hsy,Radyushkin:2017gjd} 
are in good  agreement with the observed 
\mbox{$p_3$-evolution}   patterns  obtained in lattice calculations.

In the present talk, we outline the results and  ideas formulated in   our 
next paper  \cite{Radyushkin:2017cyf}.
First, it was  demonstrated   that the connection between 
TMDs and  quasi-PDFs  is,  in fact,  a mere consequence 
of  Lorentz invariance.  Thus, it may be derived in a much simpler way than in 
 Ref. 
\cite{Radyushkin:2016hsy}. 

 Then  we  show that   the TMD/quasi-PDF connection formula
 may be rewritten in a form that allows a simple physical interpretation.
 Namely, it  tells that  
when a  hadron is moving, the parton $k_3$ momentum  may be treated as 
coming  from two sources.   First, there is the motion
of the hadron as a whole. It contributes   
the  $xp_3$  part to the total $k_3$ value, and is 
 governed by the dependence of
 the TMD  ${\cal F} (x, \kappa^2)$ on its  first, i.e., $x$,  argument.
  The remaining part $(k_3-xp_3)$   comes from the rest-frame momentum distribution, and 
is governed by the  dependence of the TMD  on its second argument,
$\kappa^2$. 
Thus, the 
 quasi-PDFs may be treated as  hybrids of PDFs and 
momentum distributions of partons in a hadron at rest.

 Since $x$ appears in both arguments of the 
TMD, the   quasi-PDFs  have a   convolution  nature. 
This fact explains  a rather complicated pattern of the change of quasi-PDFs with 
the probing momentum $p_3$, i.e., a strong nonperturbative 
 \mbox{$p_3$-evolution.} 
 One needs to have 
rather   large values $p_3 \sim 3$ GeV to ``stop'' the 
nonperturbative 
evolution and get sufficiently close 
 to the PDF limit.

It should be emphasized that PDFs are given by the 
$k_\perp$ integral of the TMDs. Since our goal is to extract PDFs,
information about a particular shape of the $k_\perp$-dependence
is redundant. In a sense, one would prefer a situation when this 
$k_\perp$-dependence is given by a delta-function $\delta (k_\perp^2)$.
Then the quasi-PDF $Q(y,p_3)$  would coincide with the PDF $f(y)$
for all probing momenta $p_3$. However, a physical TMD 
is a more involved  function of $k_\perp$. 
What is worse,  this irrelevant  \mbox{$k_\perp$-dependence } 
of the TMDs results in a complicated structure of quasi-PDFs,
necessitating large values of $p_3$ just  {\it to wipe out} information
about the $k_\perp$-dependence. One may ask if there are 
more economical ways of eliminating the unwanted $k_\perp$ effects. 

The problem is that in TMD-based momentum representation,
quasi-PDFs are given by a convolution of PDF-type $x$-dependence 
and $k_\perp$-dependence.
The latter  is related to the momentum distribution of the hadron at rest
and basically reflects the finite size of the system.
So, our next idea in Ref. 
 \cite{Radyushkin:2017cyf}
 is that the deconvolution of  the finite-size effects 
 is much simpler in the coordinate representation. 
 To this end, we introduce the functions  $ {\cal P} (x, -z^2) $  that we call
{\it pseudo-PDFs}.  They  generalize 
the light-cone PDFs $f(x)$ onto spacelike intervals. In particular, one can take    
 $z=(0,0,0,z_3)$.   The $x$-dependence of 
 the pseudo-PDFs is obtained through  Fourier transforms 
 of the  \mbox{{\it Ioffe-time}  \cite{Ioffe:1969kf}}  {\it distributions}  (ITDs) \cite{Braun:1994jq}
 ${\cal M} (\nu, z_3^2)$ with respect to $\nu=-(pz)$. 
 It should be noted that the rest-frame momentum distribution is determined 
  by  ${\cal M} (0, z_3^2)$. 
 
  The ITDs  
are  basically given by  generic matrix elements  like $ M(z,p)= \langle p |   \phi(0) \phi (z)|p \rangle $
which are the starting point  of any lattice calculation. 
To have the ITD formulation, we should treat  $M(z,p)$
as  functions of $\nu = - (pz)$ and $z^2$ (or $\nu = p_3 z_3 $ and $z_3^2$
if we take  $z=(0,0,0,z_3)$).  
The \mbox{large-$z_3$}  behavior of the  pseudo-PDFs  is governed by the same  nonperturbative 
physics that determines the $k_\perp$-dependence of   TMDs.
To get PDFs, one should either take small $z_3$ directly,
or extrapolate $ {\cal P} (x, z_3^2) $ to small $z_3$ values.
In this sense, taking small $z_3$ for pseudo-PDFs is analogous 
to taking large $p_3$ for quasi-PDFs. 

However, a serious  advantage of  the pseudo-PDFs is that,  unlike the quasi-PDFs, 
they have the ``canonical''  $-1 \leq x  \leq 1$ support
 for all $z_3^2$. 
 To access  the $z_3 \to 0$ limit through  extrapolation,
 we propose to use the reduced pseudo-PDF
 \mbox{${\mathfrak P} (x, z_3^2)\equiv {\cal P} (x, z_3^2)/{\cal M} (0, z_3^2)$,} 
 in which the nonperturbative  effects due to the rest-frame density are divided out. 
 Thus, we argue that one should use the {\it reduced ITD}  
 ${\mathfrak M} (\nu, z_3^2)\equiv {\cal M} (\nu, z_3^2)/{\cal M} (0, z_3^2)$
 as the starting object for lattice calculations of PDFs. 
When  $z_3\to 0$,  the reduced ITDs obey  
the perturbative evolution  equation,
 with $1/z_3$ serving as an evolution scale parameter.

\section{Parton  Distributions}

  \subsection{Ioffe-time distributions and  Pseudo-PDFs}

%\begin{itemize}

 %\item 
 Studying hard processes, experimentalists  work with hadrons.
 %\item  
 Theorists  work with quarks. Thus,  an important object 
 is the  amplitude $T(k,p)$ describing hadron-parton transition,
 with $p$ being the  hadron momentum, and $k$ that of  the quark.
 The transition can be described  also using    the  coordinate   space
 for quarks. Then we deal with the matrix element of a bilocal operator.
 We will  write it in a generic form  $\langle p |   \phi(0) \phi (z)|p \rangle \equiv M(z,p)$ 
 using scalar fields notations  for quarks, since the basic 
concept of the parton distributions  is  not   changed  by    spin complications.

By    Lorentz invariance, the function $M(z,p)$ 
  depends on $z$ through  two scalar  invariants,
  the {\it Ioffe time}
\cite{Ioffe:1969kf}   \mbox{$(pz) \equiv -  \nu$}  
 and   the interval $z^2$
(or $-z^2$ if we want a positive value  for spacelike $z$):
\begin{align}
 M(z,p)
=  & {\cal M} (-(pz), -z^2)  \  . 
 \label{lorentz}
\end{align} 
The function 
    ${\cal M}(\nu , -z^2) $  is    the {\it Ioffe-time distribution} (ITD)  \cite{Braun:1994jq}. 

 \begin{figure}
  \centerline{\includegraphics[height=4.5cm]{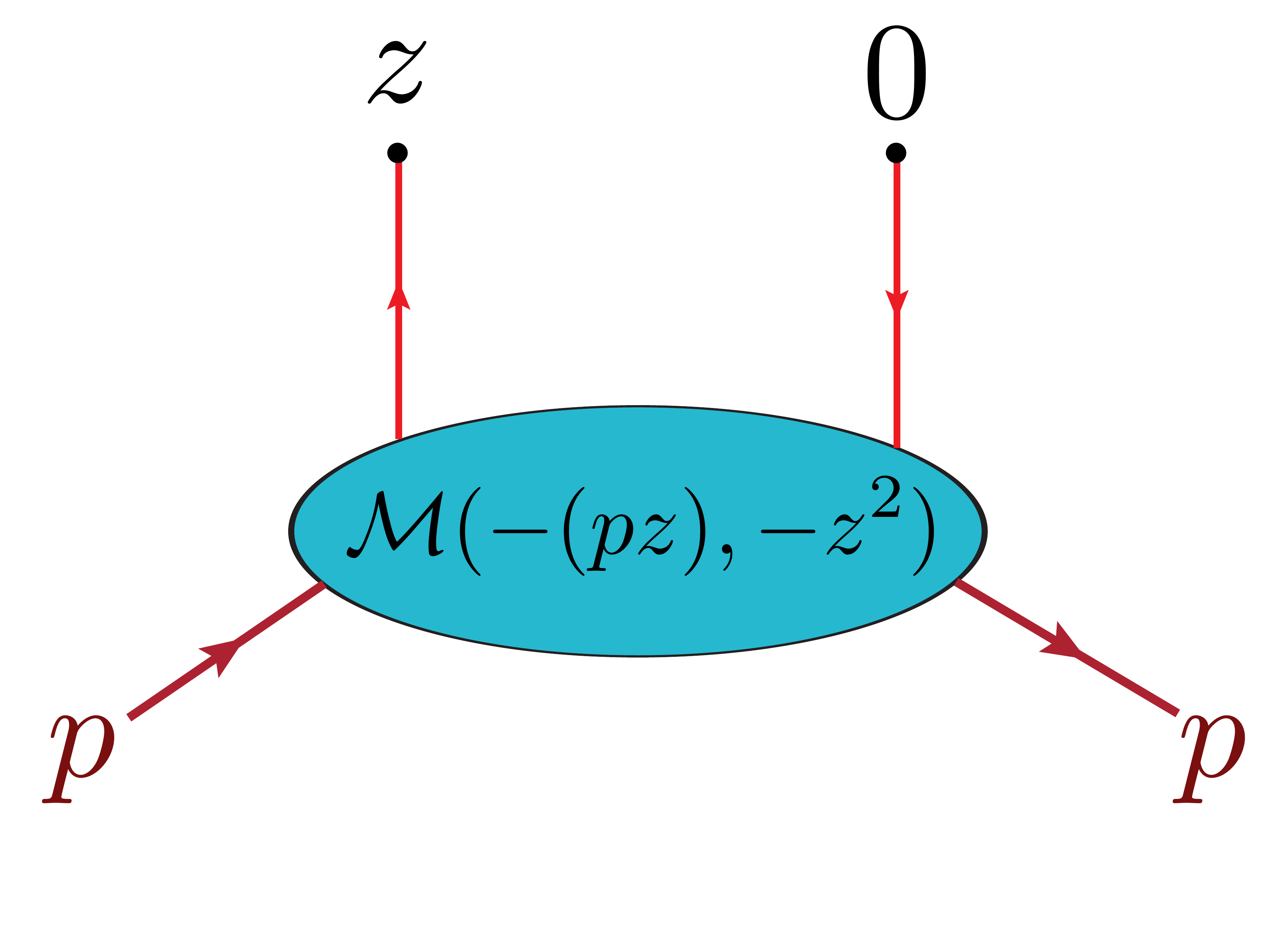}}
  \caption{Ioffe-time distribution.}
\end{figure}

 It can be shown \cite{Radyushkin:2016hsy,Radyushkin:1983wh}   
  that, for all contributing  Feynman diagrams,   
 the   Fourier transform of   $ {\cal M} (\nu, -z^2) $ with respect to $(pz)$ 
 has the $-1 \leq x \leq 1$ support, i.e., 
   \begin{align}
 {\cal M} (\nu, -z^2) 
&   = 
 \int_{-1}^1 dx 
 \, e^{-i x \nu } \,  {\cal P} (x, -z^2)  \   .
  \label{MPD}
\end{align}   
  Note that  Eq. (\ref{MPD})   gives a covariant definition of $x$.  
   There is no need to assume that 
$p^2=0$ or $z^2=0$  or take an infinite momentum frame, etc., to define $x$. 
As we will see, the function  ${\cal P} (x, -z^2)$  generalizes
the concept of the usual (or light-cone) parton distributions 
onto the case of non-lightlike intervals $z$.  Following Ref. \cite{Radyushkin:2017cyf},
we will call it 
pseudo-PDF.  
The  $-1 \leq x \leq 1$ support  region 
for  pseudo-PDF  is dictated by  analytic properties   of Feynman  diagrams, 
and 
 is  determined   
 by   the structure of denominators of propagators. 
It is not affected by 
 numerators present in non-scalar theories.  
 The inverse transformation 
   \begin{align}
     {\cal P} (x, -z^2) 
&   = \frac1{2\pi} \, 
 \int_{-\infty}^\infty  d\nu 
 \, e^{-i x \nu } \,   {\cal M} (\nu, -z^2) 
 \label{psM}
\end{align}  
may be treated as a direct definition of pseudo-PDFs as Fourier transforms 
of the ITDs  $ {\cal M} (\nu, -z^2)$   with  respect to $\nu$ for fixed $z^2$. 
Thus, pseudo-PDFs stay  just one step from the starting matrix element 
$M(z,p)$ (written  in the form of ITD), and provide the most general object from which other parton distributions
may be obtained as particular cases.

%\end{itemize}

  \subsection{Collinear Parton  Distributions,  Quasi-PDFs   and TMDs}

 Take   a  light-like $z$, say, that having just  $z_-$  component.
 Then $\nu = - p_+ z_-$, and 
 we can define the usual 
collinear (or light-cone) parton distribution $f(x)= {\cal P} (x, 0) $
  \begin{align}
  {\cal M} (-p_+ z_- , 0)  =
   \int_{-1}^1 dx \, f(x) \, 
e^{-ixp_+ z_-} \,  . 
\end{align}
It has the 
 usual  interpretation of the probability  that the parton carries fraction $x$ of the hadron's   $p_+$ 
 momentum. 
Note that the 
 $z^2 \to 0$ limit is  nontrivial in    
 QCD and other renormalizable theories, since 
 $ {\cal M} (\nu, z^2)  $ 
 has   $\sim \ln z^2$  singularities. 
The latter 
 reflect 
 perturbative evolution of parton densities. 
Within  the operator product expansion approach 
(OPE),   the   $\ln z^2$  singularities
are subtracted,    e.g.,   by dimensional renormalization, and then  $\ln (1/ z^2)  \to \ln \mu^2$.  
Resulting PDFs   depend on renormalization scale $\mu$,  
$f(x) \to   f(x, \mu^2)$.   
 If one   keeps $z^2$  spacelike, then  no  subtractions are  needed. 
 For  pseudo-PDFs  ${\cal P} (x, -z^2)$,   the interval $z^2$ serves as the ultraviolet  (UV) cut-off,
 and  $ -1/z^2$   is similar to the OPE scale  $\mu^2$.

Taking a spacelike  $z= \{0,0,0,z_3 \}$ in the frame,  where the  hadron 
momentum is  $p= (E, {\bf 0}_\perp, P)$,  one can  define quasi-PDFs 
\cite{Ji:2013dva}  as a Fourier transform of $M(z_3, P) $  with respect to $z_3$  
   \begin{align}
   Q (y, P) 
&   = \frac{P}{2\pi} \, 
 \int_{-\infty}^\infty  dz_3 
 \, e^{-i yPz_3 } \,   M(z_3, P) .
\end{align}    
It is instructive to rewrite this integral in terms of the Ioffe-time distribution 
\begin{align} 
  Q(y,  P)   =\frac{1}{2 \pi}  \int_{-\infty}^{\infty}  d\nu \, 
   \, e^{-i y  \nu}  \, {\cal M} (\nu,  \nu^2/P^2)    \   . 
   \label{QMnn}
\end{align}  
Unlike in the  pseudo-PDF  definition, the $\nu$-variable  appears 
in both arguments of the ITD.  We also see that $  Q(y,  P)  $  tends to 
the usual  PDF $f(y)$  in  the 
$P \to \infty$  limit,  as far as    ${\cal M} (\nu,  \nu^2/P^2) \to {\cal M} (\nu, 0)$.

The dependence of ${\cal M} (\nu,  -z^2)$ on $\nu$ governs the $x$-dependence 
of $f(x)$, i.e. the longitudinal momentum structure of the hadron,  while  its $z^2$-dependence 
is directly connected with the transverse momentum    distributions (TMDs). 
To show this,  let us introduce  TMDs. 
Take again the frame where $p= (E, {\bf 0}_\perp, P)$, and  choose 
 $z$ that has  $z_+=0$,  nonzero $z_-$  and,     
  in addition,   nonzero 
$z_\perp= \{z_1,z_2\}$ components.   Then    $z^2=-z_\perp^2$, and 
the  TMD  is  defined by 
     \begin{align}
 {\cal M} (\nu,  z_\perp^2) 
&   = 
 \int_{-1}^1 dx \,\, e^{i x  \nu  }  
    \int  {d^2 k_\perp }      \,  e^{-i( k_\perp  z_\perp)} 
      {\cal F} (x, k_\perp^2)  \  . 
      \label{TMD} 
\end{align}    
The
parton  again carries $xp_+$, but it also  has transverse momentum $k_\perp$
which is Fourier-conjugate to $z_\perp$. Thus, the transverse momentum
dependence of TMDs  is governed by the $z^2$-dependence of ITDs. 
 Note that, due to the rotational 
invariance in $z_\perp$ plane,   this   TMD  depends on $k_\perp^2$  only. 

While the quasi-PDF is derived from a matrix element involving  
 purely ``longitudinal''  $z=z_3$,  the dependence of 
${\cal M} (\nu,  z_3^2)$ on $z_3^2$ is given by the same function that defines 
the TMD by Eq. ({\ref{TMD}).  To relate
quasi-PDFs  and TMDs, we take $z_\perp=\{0,\nu/P \}$ in Eq.  (\ref{TMD})
and substitute the resulting representation into the expression (\ref{QMnn}) for the quasi-PDF. 
This gives \cite{Radyushkin:2016hsy,Radyushkin:2017cyf}
\begin{align}
 Q(y, P) =  & \,P 
\int_{-1} ^ {1} d  x \,  \int_{-\infty}^{\infty} d  k_1  {\cal F} (x, k_1^2+(y-x)^2P^2 ) \ . 
\label{QTMDrel}
 \end{align} 
 According to this relation, the quasi-PDF variable $y$ has the  $-\infty < y <\infty$ support,
 because the components  of the transverse momentum  $k_\perp$  in ${\cal F} (x, k_\perp^2 )$
 are  not  restricted.

 \section{Structure of Quasi-PDFs}
 
  \subsection{Momentum Distributions}

  Since  the variable $k_1$ is  integrated over in Eq. (\ref{QTMDrel}), it makes sense to introduce the function
   \begin{align}
 {\cal R} (x, k_3  )   \equiv   & \,\int_{-\infty}^{\infty} d  k_1
  {\cal F} (x, k_1^2+k_3^2)
  \  
 \label{calD}
\end{align} 
depending on the remaining momentum variable  $k_3$  only
(of course, according to Eq. (\ref{calD}),  ${\cal R} (x, k_3  ) $ depends on $k_3$ through $k_3^2$). 
 Also, instead of 
 the quasi-PDFs $Q(y,P)$  that refer to  the fraction $y \equiv k_3/P$,  
 one may    consider  distributions  in the momentum  $k_3$ itself:
 $R (k_3,P) \equiv Q(k_3/P,P)/P$.
 Then we can rewrite Eq. (\ref{QTMDrel}) as
   \begin{align}
 R(k_3, P)  =  &  \int_{-1}^1 dx\,   {\cal R} (x, k_3-xP)
  \  . 
 \label{RTMD}
 \end{align} 
For a hadron at rest, we have a one-dimensional function
  \begin{align}
 R(k_3, P=0)  \equiv   &\, r(k_3)=
   \int_{-1}^1 dx\,   {\cal R} (x, k_3  )
  \  ,
 \label{DTMDsmall}
\end{align} 
that  describes a primordial distribution of $k_3$ 
(or any other  component of  ${\bf k}$)  in  
a  rest-frame hadron. 
It 
  may be  directly  obtained  through   a  parameterization  
of  the  rest-frame density 
 \begin{align}
{\cal M} (0,z_3^2) =
\int_{-\infty}^{\infty}   dk_3 \, 
 r(k_3)  \,  e^{i k_3 z_3 } \,  
 \ . 
 \label{dk3}
\end{align} 

According to Eq. (\ref{DTMDsmall}),  the rest-frame momentum 
distribution $r(k_3)$ is obtained from $ {\cal R} (x, k_3  )$
by taking the $x$-integral.    
Similarly, integrating $ {\cal R} (x, k_3  ) $ over $k_3$ gives the collinear PDF
   \begin{align}
\,\int_{-\infty}^{\infty} d  k_3 {\cal R} (x, k_3  )=    & \,  \int 
 {d^2 k_\perp }      \, 
      {\cal F} (x, k_\perp^2)  =f(x)  \  . 
  \  
 \label{calDR}
\end{align}
Now we can give the following   interpretation of the formula (\ref{RTMD}). 
According to it, 
 in  a moving  hadron, the  parton momentum $k_3= xP + (k_3-xP)$
 has two parts. The  $xP$ part  
 comes
  from  the  motion of  the hadron as a whole 
  with the   probability  governed by $x$-dependence of
 ${\cal R} (x, k_3  ) $. 
% (or  TMD  ${\cal F} (x, \kappa^2)$).  
The probability  to get the remaining part $(k_3-xP)$  is 
 governed by   the  dependence of   ${\cal R} (x, k_3  ) $ 
 %(or   TMD)   
 on its second argument,
$k_3$, 
associated with    the primordial 
 rest-frame 
 momentum distribution.  

%\end{itemize}

\subsection{Factorized models for TMDs and quasi-PDFs}

Both arguments of 
 $ {\cal R} (x, k_3-xP)$  in Eq. (\ref{RTMD})   contain 
 the integration parameter $x$.   
  As a result, the shape of the momentum distributions  $R(k,P)$ 
 (and, hence, of the quasi-PDFs)   is  influenced by 
 the form   both of 
 PDFs  and  rest-frame distributions.   
 To illustrate   the ``hybrid'' nature of momentum distributions and quasi-PDFs,  
 we  will  use  
 a factorized model $ {\cal R} (x, k_3) = f(x) r (k_3)$. 
For  the ITD,   this Ansatz corresponds 
to the factorization assumption 
  \begin{align} {\cal M}^{\rm fact}  (\nu,-z^2) = {\cal M} (\nu,0){\cal M} (0,-z^2) \  .
    \end{align}
 A popular choice is a    Gaussian dependence of TMDs on
$k_\perp$. It gives 
  \begin{align}
r_G (k_3) = \frac{1}{\sqrt{\pi} \Lambda}  e^{-k_3^2/\Lambda^2} \ \  {\rm or}  \ \   
r_G (z_3^3) =   e^{-z_3^2 \Lambda^2/4 }  \ \ \  {\rm for \ \  the \ \ rest \  frame \ \ density .}  \ \ \ \  
\end{align} 
Then the  factorized Gaussian model for the momentum distribution has the form 
 \begin{align}
 R_G^{\rm fact}  (k_3, P)  = &\frac{1}{\Lambda \sqrt{\pi} }  \,
 \int_{-1}^1 dx\,  %  
f(x) \, 
  e^{- (k_3- xP)^2 / \Lambda^2 }\  . 
\end{align} 
For  PDF we choose a   simple function 
$f(x)=4(1-x)^3 \theta (0\leq x \leq 1)$    resembling valence quark distributions.  
 From Fig. \ref{R},    one can see that  the curve for $R(k,P)$ 
changes from a Gaussian shape for small $P$  to a shape resembling  a
 stretched PDF  for large $P$.  For small $P/\Lambda$ values, we may approximate
 \begin{align}
% \nonumber 
 R (k_3, P)  = & \,
 \int_{-1}^1 dx\,  f(x) \, %    
 r(k_3-xP)  \approx  r(k_3-\tilde xP)  
 \   
  \label{Dfac}
\end{align}  
($\tilde x = $  average $x$,  in our model $\tilde x =0.2$), 
i.e., for small $P$, the  $R (k_3, P) $  curve  
approximately   keeps  its shape,  but the maximum shifts to the right when $P$ increases. 
For large $P$, we have  
  \begin{align}
r_G(k_3-xP) =  & 
\frac{1}{\sqrt{\pi } \Lambda}  e^{-(k_3-xP)^2/\Lambda^2}
\to \frac1{P} \, \delta (x-k_3/P) \ , 
\end{align} 
i.e., the combination $P\, R (k_3, P)$ corresponding to   quasi-PDF
 $Q(y=k_3/P,P)$ 
 in the  large $P$ limit converts into  a scaling function  $f(k_3/P)=f(y)$
coinciding with the input PDF. 
\begin{figure}[t]
    \centerline{ \includegraphics[width=2in]{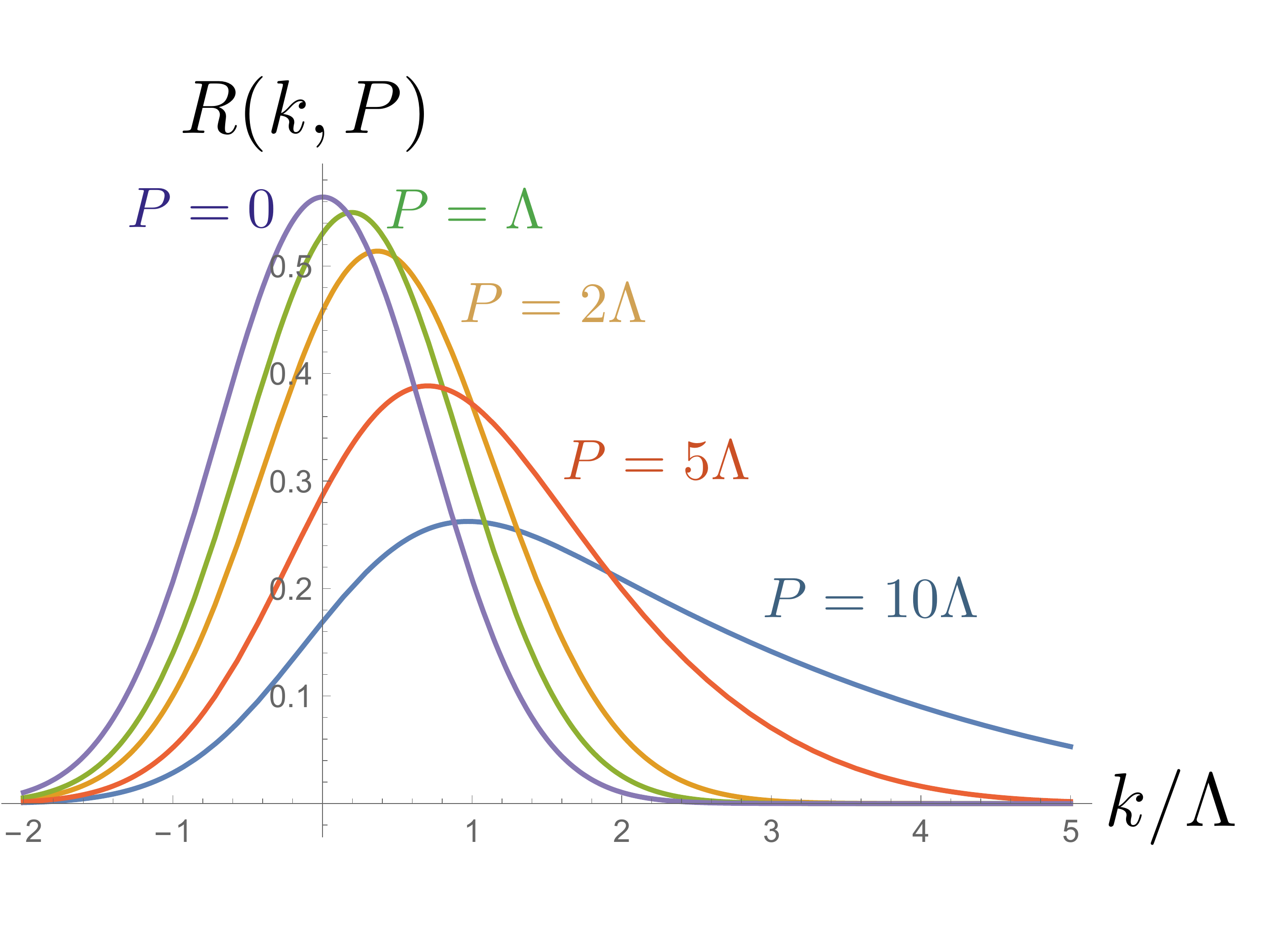}  \  \includegraphics[width=2in]{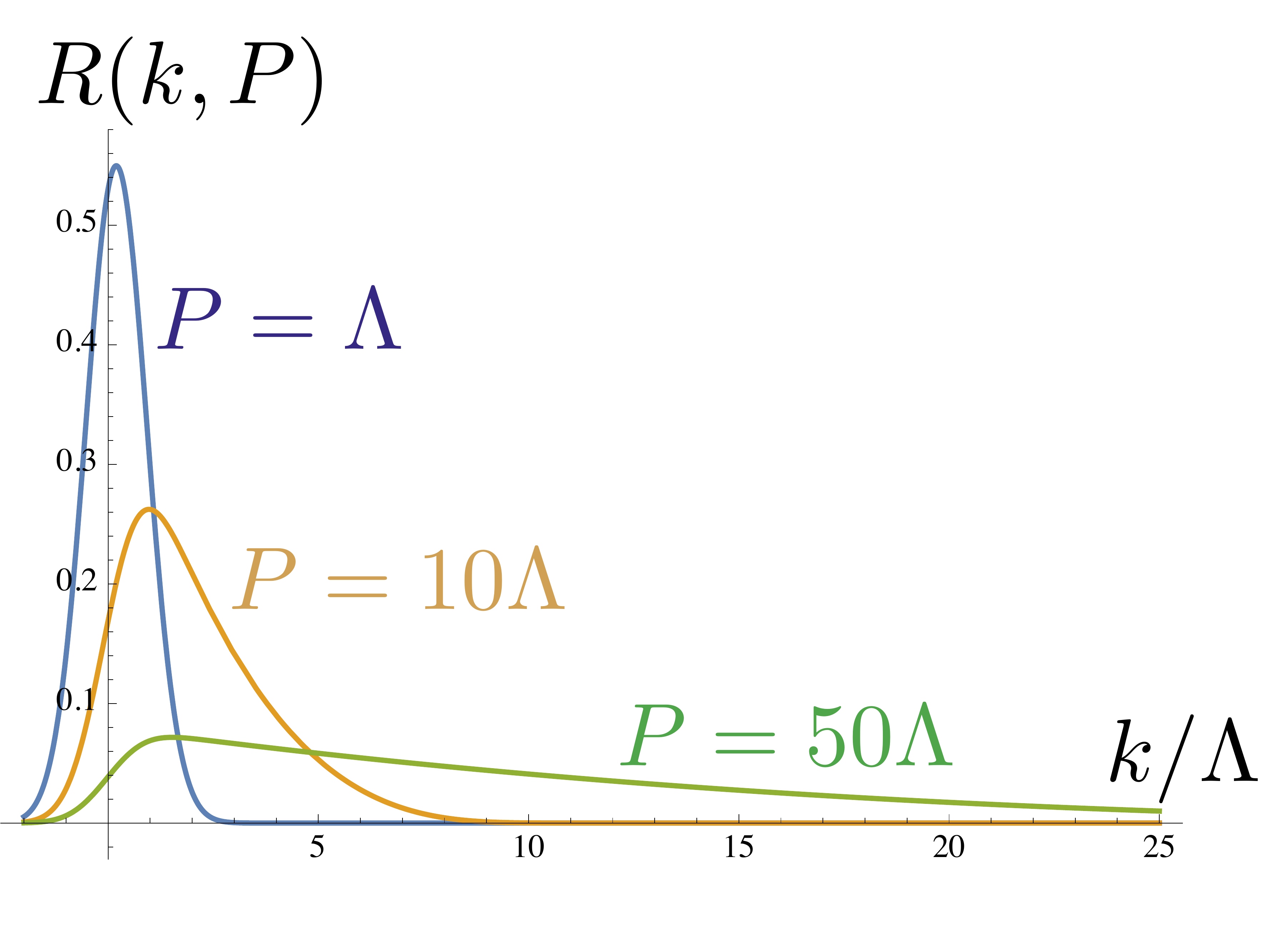}
    \includegraphics[width=2in]{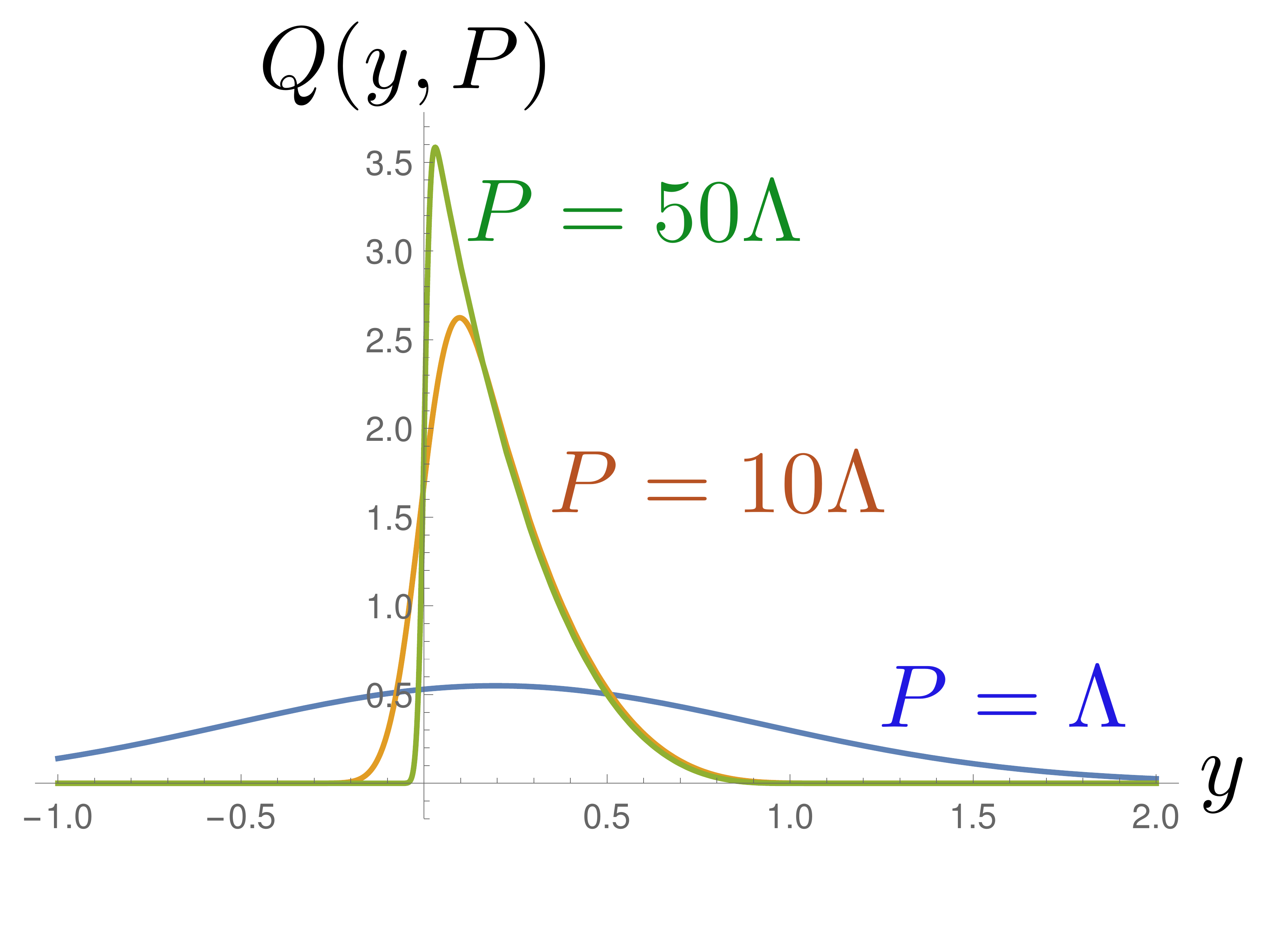}  }
    \caption{Momentum distribution $R(k,P)$ and quasi-PDF $Q(y,P)$  for different momentum $P$  values. }
    \label{R}
    \end{figure}

%\end{itemize}

%\end{itemize}

\subsection{QCD  case}

%\begin{itemize}

%\item 
  In QCD  we deal with matrix elements  of the 
   $
 {\cal M}^\alpha  (z,p) \equiv \langle  p |  \bar \psi (0) \,
 \gamma^\alpha \,  { \hat E} (0,z; A) \psi (z) | p \rangle \  
$ type, where  $
{ \hat E}(0,z; A)$ 
is the   standard  $0\to z$ straight-line gauge link. 
Due to the vector index $\alpha$, the function $ {\cal M}^\alpha  (z,p) $ may be 
decomposed into $p^\alpha$ and $z^\alpha$ parts
\begin{align}
     {\cal M}^\alpha  (z,p) =2 p^\alpha  {\cal M}_p (-(zp), -z^2) + z^\alpha  {\cal M}_z (-(zp),-z^2) \  . 
    \end{align}  
In the standard definition of the TMD, we 
have   $z_+=0$  and  take $\alpha=+$.  As a result, the  \mbox{$z^\alpha$-part} drops out, and  
TMD  ${\cal F}(x, k_\perp^2)$   
is  related to  $ {\cal M}_p (\nu, z_\perp^2)$  by  the scalar formula.  
To remove the  \mbox{$z^\alpha$-contamination}  
from  quasi- PDF, we 
 take  the  time component of   \mbox{$ {\cal M}^\alpha  (z=z_3,p)$}   and define
 \begin{align}
&  {\cal M}^0   (z_3,p)   
  = 2 p^0 
 \int_{-1}^1 dy\,  
Q(y,P) \, 
 \,  e^{i  y Pz_3 }  \  .
\end{align} 
Then 
quasi-PDF   $Q (y,P)$ is  related
to TMD  ${\cal F}(x, k_\perp^2)$ by  the scalar   formula (\ref{QTMDrel}).

%\end{itemize}

\section{Quasi-PDFs vs Pseudo-PDFs  and Ioffe-time Distributions}

\begin{figure}[b]
       \centerline{\includegraphics[width=2.5in]{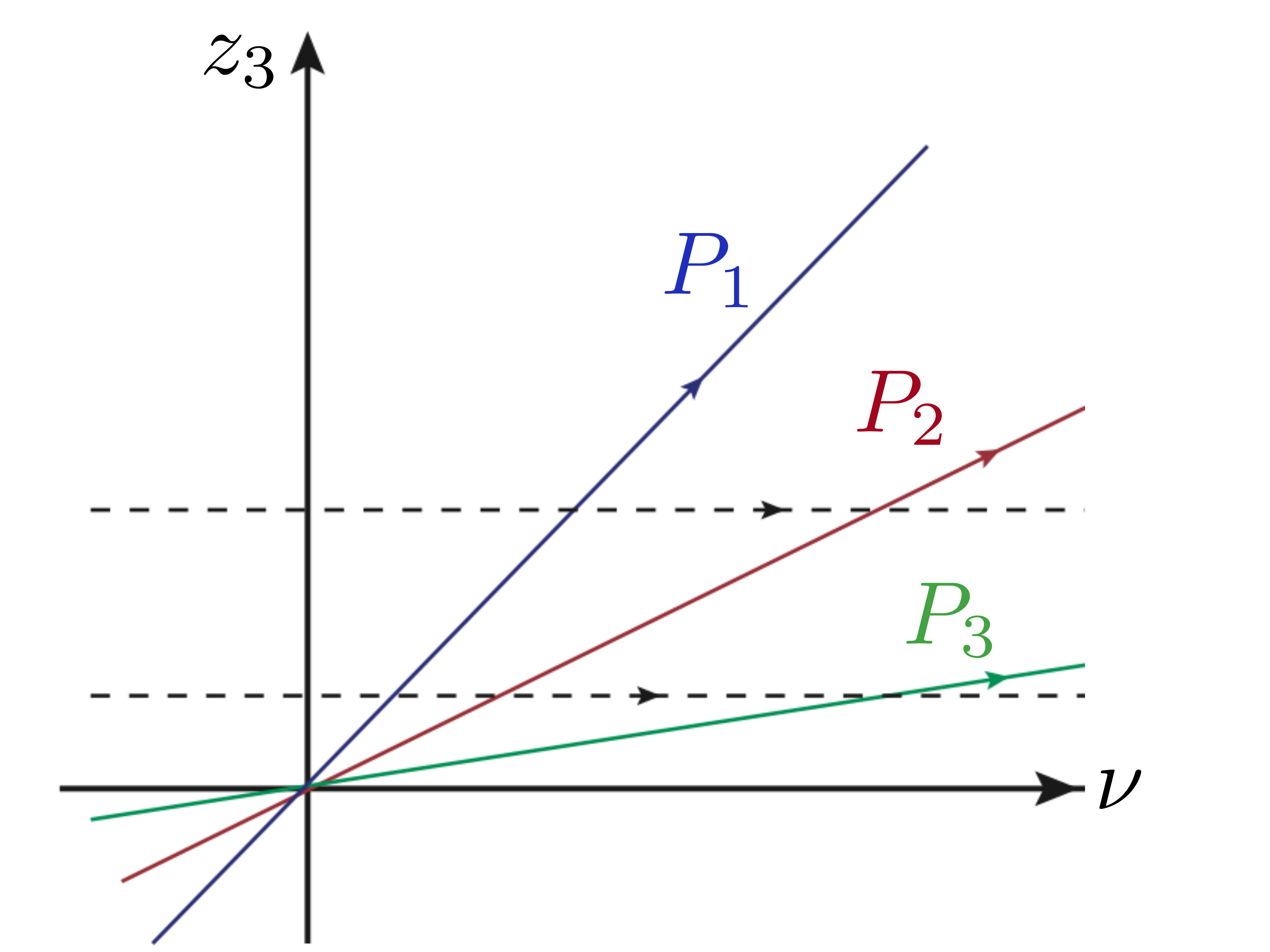}}
       \caption{Lines of integration for quasi-PDF $Q(y,P)$ in the $\{\nu,z_3 \}$ plane 
       (solid lines) and for pseudo-PDFs ${\cal P} (x, z_3^2)$ (dashed lines). }
       \label{plane}
       \end{figure}

According to the definition of quasi-PDF $Q(y,P)$  in Eq.(\ref{QMnn}),   they are obtained from the ITD
${\cal M} (\nu, z_3^2)$
by  integration   over $z_3=\nu/P$ lines
in the $\{\nu,z_3 \}$ plane, see Fig. \ref{plane}. 
They tend to the horizontal $z_3=0$ line in the $P \to \infty$ limit, and the resulting quasi-PDFs approach  PDF.  
It should be noted that  this approach is non-trivial, since 
 $Q(y,P)$ has perturbative evolution with respect to $P$ for large $P$.  
 In general, quasi-PDFs have the 
 $-\infty < y <\infty$ support region.  As we have seen, in the case of  the soft factorized  
 models, the support  shrinks  to $-1< y <1$ in the $P \to \infty$ limit. 
 If one adds perturbative corrections due to hard gluon exchanges,
 they generate terms with the  $-\infty < y <\infty$ support  even in  the $P \to \infty$ limit.
 Such terms should be removed through the use of matching conditions \cite{Ji:2013dva}. 

 Pseudo-PDFs, according to  their definition (\ref{psM}), are given by 
    integration of ${\cal M} (\nu, z_3^2)$ over $z_3=$const lines.
They always have the  $-1\leq x \leq 1$ support. 
For  small  $z_3^2$, the pseudo-PDFs 
${\cal P} (x, z_3^2)$ have  perturbative evolution with respect to $1/z_3$. 
At the leading logarithm level, they are close to usual PDFs
 $f(x, C^2/z_3^2)$  with $C$  being the matching coefficient, 
$C_{\rm \overline {MS}}= 2e^{-\gamma_E}\approx 1.12$. 

%\end{itemize}

The fact that quasi-PDFs $Q(y,P)$ are    given by  integration of ${\cal M} (\nu, z_3^2)$ over the 
$z_3=\nu/P$ lines
leads to their  
  $x$-convolution
structure,   even if  ${\cal M} (\nu, z_3^2)$   factorizes, i.e.,
 ${\cal M} (\nu, z_3^2)= {\cal M} (\nu, 0) {\cal M} (0, z_3^2)$. 
 An alternative approach \cite{Radyushkin:2017cyf} is to 
convert   lattice data for  ${\cal M} (Pz_3, z_3^2)$ into the data for   ${\cal M} (\nu, z_3^2)$.  
The next step  is to take the reduced function
 \begin{align}
{\mathfrak M} (\nu, z_3^2) \equiv \frac{ {\cal M} (\nu, z_3^2)}{{\cal M} (0, z_3^2)} \  ,
\end{align}
i.e.   divide ITD ${\cal M} (\nu, z_3^2)$ by the rest-frame density ${\cal M} (0, z_3^2)$.
In factorized case,  the reduced ITD  converts into   $ {\cal M} (\nu, 0)$,
and what formally remains is to    
 take its Fourier transform to get  PDF $f(x)$.  
Another advantage of using the reduced ITD is that the 
 $z_3^2$-dependence due to self-energy of gauge link  cancels in the ratio,
 because the UV-induced $z_3^2$-dependence is multiplicative
 (see Refs.  \cite{Ishikawa:2016znu,Ishikawa:2017faj,Ji:2017oey,Green:2017xeu} for recent progress in this field.)

%\end{itemize}

   \subsection{Evolution of Ioffe-time  distributions}

%\begin{itemize}

%\item 
Originally, the  Ioffe-time distributions $Q(\nu, \mu^2)$
were defined  \cite{Braun:1994jq} 
as functions whose Fourier transforms with respect to $\nu$ 
were given by usual OPE  PDFs $f(x,\mu^2)$.
Thus, their dependence on the renormalization parameter $\mu$
(say, $\overline{\rm MS}$ scale)  is completely determined 
by the evolution equation for PDFs  $f(x,\mu^2)$.
In case of pseudo-PDFs, the parameter $1/z_3$ for small $z_3$
plays the role of $\mu$. A subtlety is that  ${\cal M} (\nu, z_3^2)$ 
has an extra $z_3$ dependence induced by the renormalization of the gauge link.
However, this $z_3$-dependence cancels in the reduced ITD  ${\mathfrak M} (\nu, z_3^2)$.
As a result, for small $z_3^2$ we have the  
leading-order  evolution equation 
    \begin{align}
    \frac{d}{d \ln z_3^2} \,  
{\mathfrak M} (\nu, z_3^2)    &=- \frac{\alpha_s}{2\pi} \, C_F
\int_0^1  du \,   B ( u )   {\mathfrak  M} (u \nu, z_3^2)  
\nonumber  
 \end{align}
with the same 
 nonsinglet evolution kernel 
\begin{align}
B (u)    &=  \left [ \frac{1+u^2}{1- u} \right ]_+  \ 
\nonumber  
 \end{align} 
as in Ref. \cite{Braun:1994jq}.
Examples of real and imaginary parts of ITD are shown in Fig. \ref{MBM}, 
together with functions $B \otimes {\cal M}$ governing their perturbative evolution.
One can see that there are no perturbative evolution for  ${\cal M} (0, z_3^2)$ 
[vector current is conserved].  Also, ${\rm Im} \ {\cal M} (0,z_3^2)=0$, i.e., the rest-frame 
distribution $ {\cal M} (0,z_3^2)=0$ is purely real.

\begin{figure}[t]
   \centerline{\includegraphics[width=2.5in]{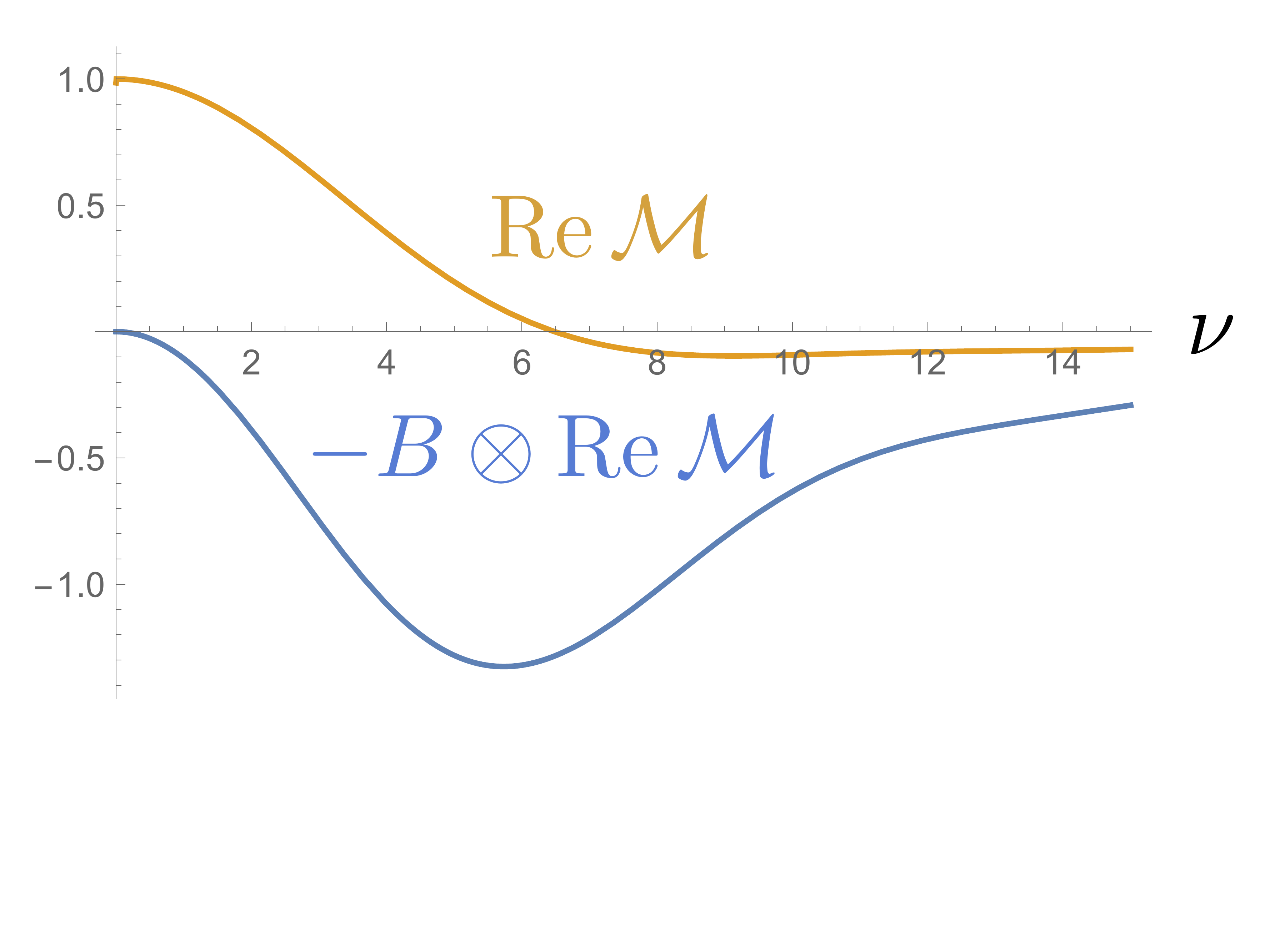} \  \   \includegraphics[width=2.5in]{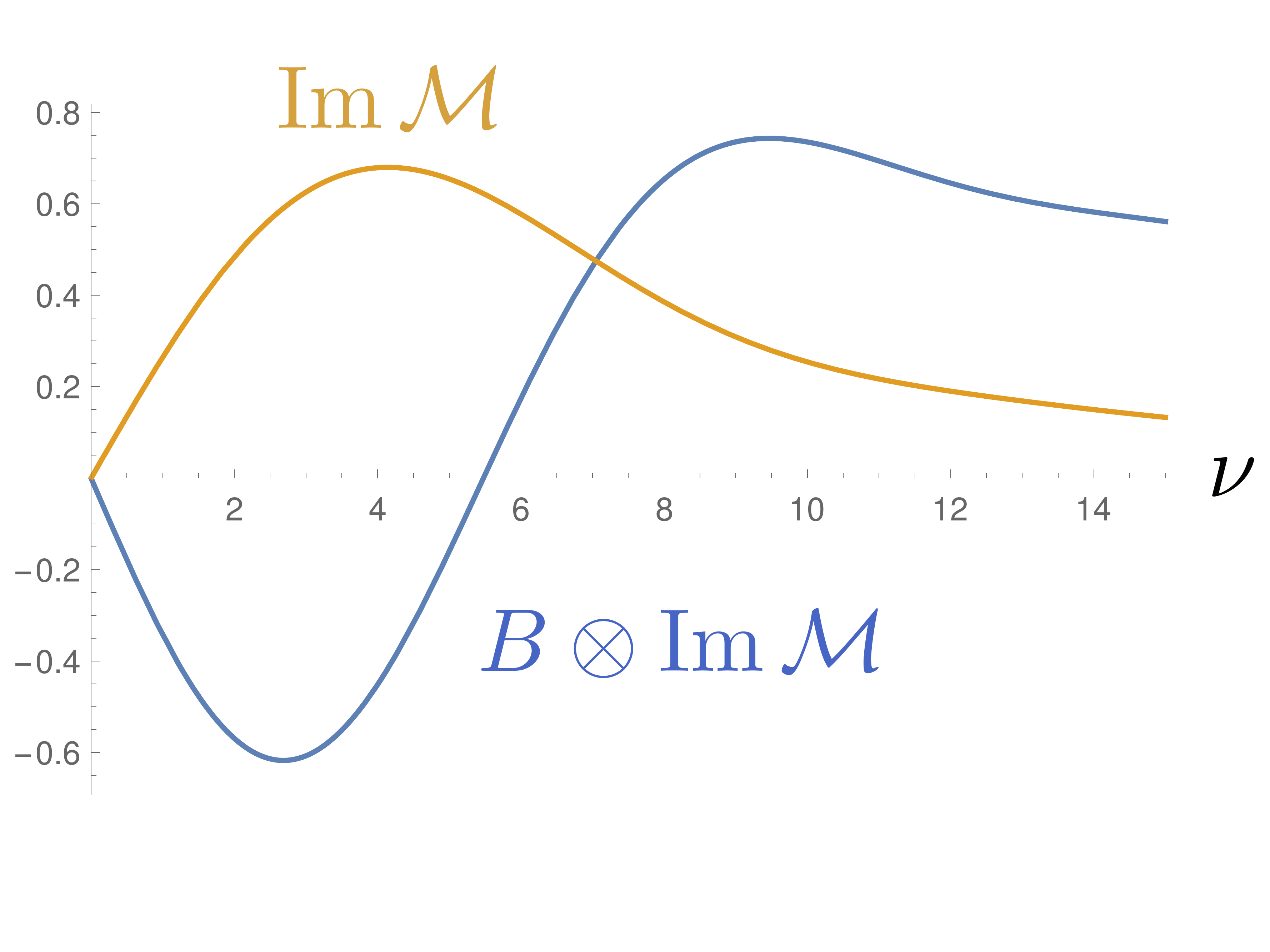}}
    \vspace{-0.6cm}
    \caption{Real and imaginary parts of model Ioffe-time distribution ${\cal M} (\nu,0)$  and
    the function $B \otimes {\cal M}$  governing their  evolution.
    \label{MBM}}
    \end{figure}

%\item 

%\end{itemize}

\subsection{Nonfactorizable  cases}

The  perturbative  evolution produces $z_3^2$-dependence for the reduced ITD  ${\mathfrak M} (\nu, z_3^2)$.
Hence, it will  unavoidably  violate  factorization.  
This $z_3^2$-dependence should be visible in the data  as 
$\ln (1/z_3^2 \Lambda^2)$  spikes for small $z_3^2$.

Take for illustration ${\cal  P}^{\rm soft}  (x, z_3^2) = f(x)e^{-z_3^2  \Lambda^2/4} $
for the soft part 
(corresponding to the TMD ${\cal F}^{\rm soft} (x, k_\perp^2)  =  f(x)  e^{-k_\perp^2  /  \Lambda^2}/\pi \Lambda^2 $) 
and choose $\alpha_s/\pi =0.1$ for the hard part in  which we use 
the incomplete gamma-function $ \Gamma[0, z_3^2 \Lambda^2/4]$
instead of a straightforward $\ln (1/z_3^2 \Lambda^2)$ function.
In such a model for the hard part, the evolution stops for large $z_3^2$.  

     For the real part of the ITD, the evolution effects are the largest 
     for $\nu \sim 6$. As one can see from Fig. \ref{evol}, they are clearly visible for 
     $z_3 \Lambda \simeq 1.5$. Assuming  $\Lambda =300{\rm MeV}$,
      we obtain that    $z_3  \Lambda$ = 1.5 for   $z_3 =1$ fm. 

\begin{figure}[b]
    \centerline{\includegraphics[width=3in]{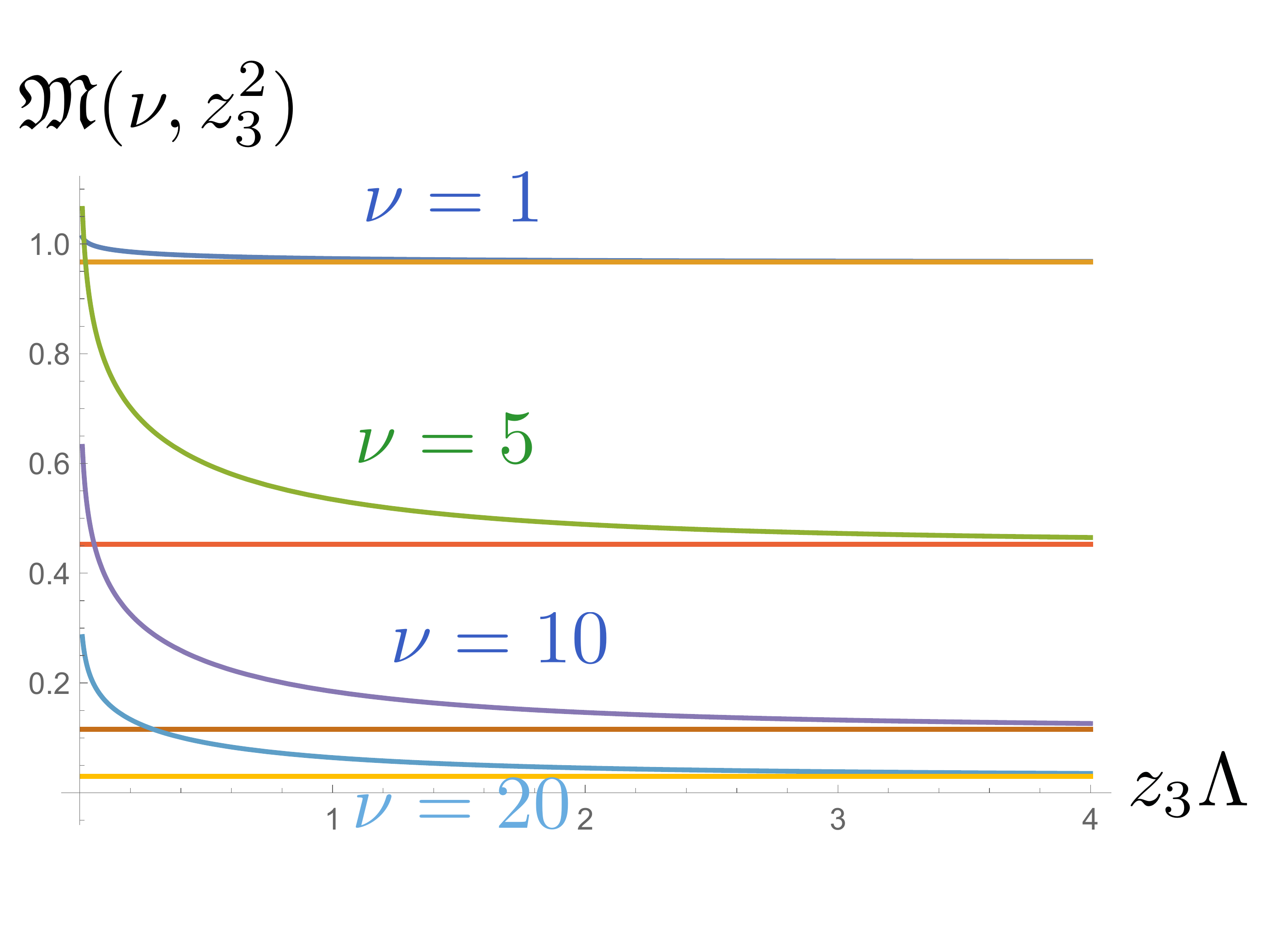} \ \ \includegraphics[width=3in]{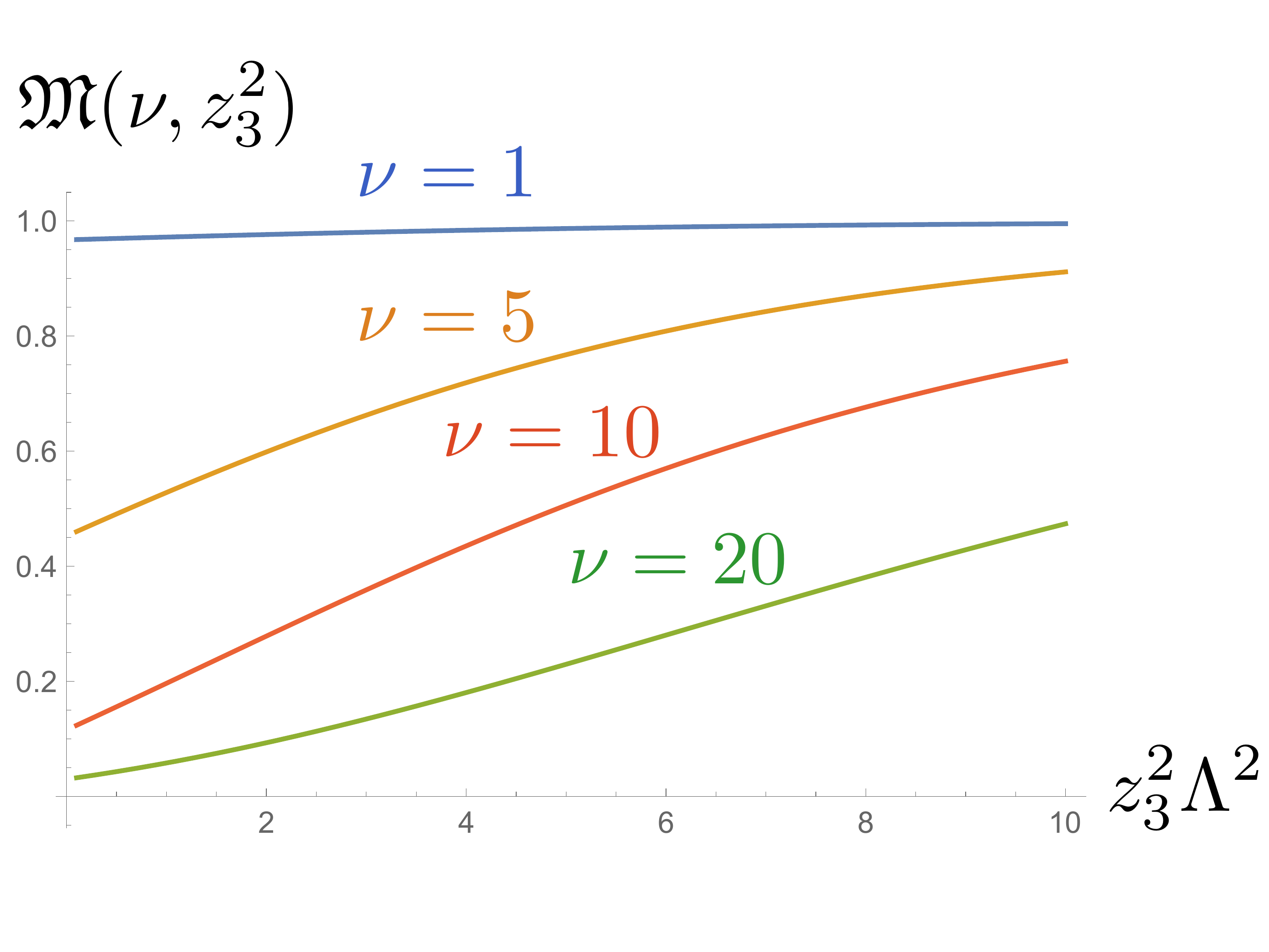}}
    \vspace{-0.6cm}
    \caption{Violation of factorization for the reduced ITD due to perturbative evolution (left)
    and due to non-factorizable  form of TMD (right). }
    \label{evol}
    \end{figure}

The reduced ITD   
${\mathfrak M} (\nu, z_3^2)$   may also have 
  residual \mbox{$z_3^2$-dependence}  from the 
 violation of factorization in the   soft part. 
To illustrate these effects, we take  the model pseudo-PDF ${\cal  P}^{\rm soft}  (x, z_3^2) = f(x)e^{-x (1-x) z_3^2  \tilde \Lambda^2/4} $
corresponding to the model TMD 
 ${\cal F}^{\rm soft} (x, k_\perp^2)  =  f(x)  e^{-k_\perp^2 /x (1-x)   \tilde \Lambda^2} /[\pi x (1-x)  \tilde \Lambda^2]$),
 whose  dependence on $k_\perp^2$  comes through the $k_\perp^2 /x (1-x) $ combination
 advocated by the light-front quantization proponents.  
 To have the same  $\langle k_\perp^2 \rangle $ we need   $\tilde \Lambda^2  =\frac{15}{2} \Lambda^2$. 
 The $z_3=1$ fm distance on the right graph of Fig. \ref{evol}  corresponds to 
  $z_3^2  \Lambda^2$ = 2.25.

In a  situation when factorization is violated  both by evolution 
and non-perturbative effects, a possible
  strategy is  to extrapolate 
${\mathfrak M} (\nu, z_3^2)$ to $z_3^2=0$ from not too small
values of $z_3^2$, say,  from those above 0.5 fm$^2$.
 The resulting function  ${\cal M} ^{\rm soft}(\nu,0)$
may be treated as 
the Ioffe-time distribution producing the PDF $f_0(x)$ 
 ``at low normalization point''.  The remaining 
$\ln (1/z_3^2 \Lambda^2)$  spikes  at small $z_3$ will generate its evolution.

%\item    

%\end{itemize}

  \section{Summary}

In this talk, we described our  recent work 
\cite{Radyushkin:2016hsy,Radyushkin:2017cyf} 
on the structure of parton quasi-distributions. 
We found that the quasi-PDFs are  { hybrids} of PDFs and 
primordial rest-frame 
momentum distributions.  
The resulting complicated {convolution  nature} of quasi-PDFs 
necessitates   large probing momenta  $p_3 \gtrsim $ \mbox{ 3 GeV}  to wipe out  the 
primordial  effects.

  To avoid convolution structures, we proposed 
    to use 
{pseudo-PDFs}   $ {\cal P} (x, z_3^2)$, the functions   most closely 
  related  (by  a  Fourier transform)  to  the 
 {Ioffe-time  distributions}   ${\cal M} (\nu, z_3^2) $,
 the primary objects both for continuum and lattice studies of parton distributions. 
 One of the advantages of the 
pseudo-PDFs  is that they have  the 
 same ``canonical'' {$-1 \leq x  \leq 1$}  support  as usual PDFs.  
 Furthermore, their 
 $z_3^2$-dependence  for small $z_3^2$  is governed 
 by  a {usual  evolution equation}. 
 
  An important ingredient of the 
 proposed program for  the pseudo-PDF-based  lattice extraction of PDFs 
 is the 
use of the reduced     Ioffe-time  distributions  given by the {ratio ${\cal M} (\nu, z_3^2)/{\cal M} (0, z_3^2)$ } 
 in which the   $z_3^2$-dependence of the primordial
rest-frame density   ${\cal M} (0, z_3^2)$  is divided out 
from the original Ioffe-time distribution  ${\cal M} (\nu, z_3^2)$.
The use of this 
ratio also provides a very simple and efficient way for 
getting rid of the     \mbox{ $z_3^2$-dependence}   related to the 
ultraviolet divergences generated by the self-energy and vertex corrections
to the gauge  link.  
 
While the write-up of this talk was in preparation, the  exploratory 
lattice studies of the pseudo-PDFs  have been performed   \cite{Orginos:2017wcl,Karpie:2017bzm} .
Their results have completely confirmed expectations formulated in Ref. \cite{Radyushkin:2017cyf}.

%%%%%%%%

\acknowledgments 

I thank  V. M.  Braun,  X. D. Ji  and J.-W. Qiu for discussions
 and suggestions.
  I am especially
grateful to \mbox{K. Orginos}  for stimulating discussions and suggestions concerning 
 the lattice implementation of the  approach. 
This work is supported by Jefferson Science Associates,
 LLC under  U.S. DOE Contract \#DE-AC05-06OR23177
 and by U.S. DOE Grant  \mbox{\#DE-FG02-97ER41028. }
 
 \providecommand{\href}[2]{#2}\begingroup\raggedright\endgroup

    \end{document}